\documentclass[10pt,english]{article}
\usepackage{times}
\usepackage[T1]{fontenc}
\usepackage[latin1]{inputenc}
\usepackage{amsmath}
\usepackage{amssymb}

\makeatletter

\usepackage{slashed}

\usepackage{babel}
\makeatother
\begin{document}

\title{\textbf{Dimension-five effective operators in electro-weak $SU(4)_{L}\otimes U(1)_{X}$
gauge models}}

\author{ADRIAN PALCU}

\date{\emph{Faculty of Exact Sciences - {}``Aurel Vlaicu'' University
Arad, Str. Elena Dr\u{a}goi 2, Arad - 310330, Romania}}

\maketitle
\begin{abstract}
We prove in this paper that the electro-weak \textbf{$SU(4)_{L}\otimes U(1)_{X}$}
gauge models with spontaneous symmetry breaking can offer a natural
framework for generating neutrino masses by simply exploiting the
tree level realization of dimension-five effective operators. The
novelty of our approach resides in the fact that the scalar sector
needs not to be enlarged, since these operators are constructed as
direct products among scalar multiplets already existing in the model.
There is a unique generic matrix for Youkawa couplings in the neutrino
sector. The charged leptons are already in their diagonal basis. This
framework can lead to a suitable fit of the established phenomenology
for the left-handed neutrinos, while the right-handed neutrino masses
come out in the sub-$keV$ region, independently of the cut-off $\Lambda$.
The latter introduces in the theory an intermediate scale (however,
more close to GUT than to SM) at about $10^{12}GeV$ which is a crucial
ingredient for the left-handed neutrino phenomenology. 


\end{abstract}

\section{Introduction}

One of the main challenges in particle physics \cite{key-1} today
is the neutrino mass issue. Both its origin and order of magnitude
are still awaiting a compelling theoretical explanation. Observational
collaborations \cite{key-2} such as SuperKamiokande, K2K, SNO, KamLAND,
LSND and others have stated the phenomenon of neutrino oscillation
as an indisputable evidence. This state of affairs claims for tiny
but non-zero masses for neutrinos, regardless they will prove themselves
as Dirac or Majorana particles. Consequently, the theory is called
to supply a convenient framework and suitable mechanisms for generating
these tiny masses, assuming that the Standard Model (SM) does not
necessarily include right-handed neutrinos (otherwise unavoidable
ingredients in accomplishing a non-zero mass term). The experimental
side of the neutrino mass issue enforces certain restrictions, namely
the observed mass splitting ratio $r_{\Delta}=\Delta m_{\odot}^{2}/\Delta m_{atm}^{2}\simeq0.033$
and particular patterns for the mixing angles $\theta_{\odot}\simeq34^{\circ}$
and $\theta_{\odot}\simeq45^{\circ}$, along with a likely $\theta_{13}\simeq0$.
The absolute mass hierarchy remains still undetermined on theoretical
grounds. What we only know at present is that it lies in the sub-$eV$
region\cite{key-1}.

We mention the see-saw mechanism\cite{key-3} (with its variants)
and higher-dimension operators in effective theories \cite{key-4}
among the most appealing theoretical devices designed to accommodate
neutrino phenomenology and predict viable consequences of it at low
energies. These approaches generally require a larger framework than
the one offered by the SM. For instance, the canonical seesaw mechanism
essentially relies on a new higher scale (not subject in the gauge
symmetry of the SM, but violating $B-L$ symmetry) to generate Majorana
right-handed masses. In a self-explanatory notation, the seesaw $6\times6$
matrix can be put as:

\begin{equation}
M^{M+D}=\left(\begin{array}{ccc}
0 &  & m_{D}^{T}\\
\\m_{D} &  & M_{R}\end{array}\right).\label{Eq. 1}\end{equation}

By diagonalizing it one can get both the left-handed and right-handed
neutrino masses as

\begin{equation}
M\left(\nu_{L}\right)=-m_{D}^{T}M_{R}^{-1}m_{D}\quad,\quad M\left(\nu_{R}\right)=M_{R}.\label{Eq. 2}\end{equation}

Therefore it seems worthwhile to investigate some extensions of the
SM that include in a natural way right-handed neutrinos. During the
last two decades, gauge models such as $SU(3)_{C}\otimes SU(3)_{L}\otimes U(1)_{X}$
\cite{key-5,key-6} ) and $SU(3)_{C}\otimes SU(4)_{L}\otimes U(1)_{X}$
\cite{key-7} - \cite{key-9} have been intensely investigated. They
offer promising results when addressing the neutrino issue. In most
of the 3-3-1 models \cite{key-6} a scalar sextet must be added to
the Higgs sector in order to give rise to some seesaw mechanisms.
In 3-4-1 models the same strategy was considered in Refs. \cite{key-8}
where a new Higgs decaplet is introduced in the scalar sector of the
model in order to set up proper seesaw terms. Yet, this can affect
the boson masses previously calculated, while some \emph{ad hoc} hypothesis
on the breaking scales of the decaplet must be speculated. 

Our approach here avoids such new ingredients and proves itself able
to set up the canonical seesaw mechanism via dimension-five effective
operators constructed out of the existing ingredients in the model.
It deals with the class of 3-4-1 models without exotic electric charges,
so that even the exotic new quarks exhibit the $\pm\frac{2}{3}$ and
$\pm\frac{1}{3}$ electric charges and mix with the traditional quarks
of the SM. It also does not involve any enlargement of the scalar
sector, but just makes use of the cut-off $\Lambda$ up to which the
model works as a safe renormalizable effective theory. The latter
not only determines the highest bound for the validity of the theory
but also play a crucial role in predictions regarding the order of
magnitude of the active neutrinos' masses.

Our paper is conceived as follows: after a brief overview of the gauge
model and its main phenomenological features in Sec.2, we focus on
its lepton content in Sec. 3 presenting the mass generating procedure
based on dimension-5 effective operators in the neutrino sector. Sec.
4 is reserved for some numerical estimates and predictions in a particular
scenarios taken into consideration, while in Sec. 5 we sketch our
conclusions.

\section{Electro-weak $SU(4)_{L}\otimes U(1)_{X}$ gauge model}

We start here by briefly presenting the particular 3-4-1 gauge model
under consideration here. However, the reader can find its phenomenological
details treated \emph{in extenso} in Refs. \cite{key-8}. Evidently,
our focus will go to its lepton sector, as we intend to exploit the
realization of dimension-five effective operators responsible for
giving rise (at the tree level) to the well known seesaw terms. 

In the electro-weak gauge group ($SU(3)_{C}\otimes SU(4)_{L}\otimes U(1)_{X}$
the electric charge operator is a linear combination of diagonal Hermitian
generators from Cartan sub-algebra. It is realized in the manner:
$Q=T_{3L}+\frac{1}{\sqrt{3}}T_{8L}+\frac{1}{\sqrt{6}}T_{15L}+\frac{1}{2}XI$
where $T_{a}=\frac{1}{2}\lambda_{a}$ normalized as $Tr(T_{a}T_{b})=\frac{1}{2}\delta_{ab}$. 

The lepton representations in this model are:

\begin{equation}
f_{L}=\left(\begin{array}{c}
e\\
\nu_{e}\\
N_{e}\\
N_{e}^{\prime}\end{array}\right)_{L},\quad\left(\begin{array}{c}
\mu\\
\nu_{\mu}\\
N_{\mu}\\
N_{\mu}^{\prime}\end{array}\right)_{L},\quad\left(\begin{array}{c}
\tau\\
\nu_{\tau}\\
N_{\tau}\\
N_{\tau}^{\prime}\end{array}\right)_{L}\sim(\mathbf{1,4^{*}},-1/2)\label{Eq. 3}\end{equation}

\begin{equation}
e_{R},\mu_{R},\tau_{R}\sim(\mathbf{1,1},-2)\label{Eq. 4}\end{equation}

For the sake of completeness, we present, in addition, the quark representations
of the 3-4-1 model of interest here. In order to cancel all the chiral
anomalies, two left-handed quark families must transform as $Q{}_{\alpha}=\left(\begin{array}{cccc}
u_{\alpha} & d_{\alpha} & D_{\alpha} & D_{\alpha}^{\prime}\end{array}\right)_{L}^{T}\sim(\mathbf{3},\mathbf{4},-1/6)$, differently from a third generation which does it in the manner
$Q{}_{3}=\left(\begin{array}{cccc}
d_{3} & u_{3} & U & U^{\prime}\end{array}\right)_{L}^{T}\sim(\mathbf{3},\mathbf{4^{*}},5/6)$. Their right-handed partners are singlets with respect to the electro-weak
gauge group, namely $d_{3R}$, $d_{\alpha R}$, $D_{\alpha R}{}$,
$D_{\alpha R}^{\prime}$$\sim(\mathbf{3},\mathbf{1},-2/3)$, $u_{3R}$,
$u_{\alpha R}$, $U_{R}$, $U_{R}^{\prime}$$\sim(\mathbf{3},\mathbf{1},4/3)$,
with $\alpha=1,2$. Capital letters denote exotic quarks, yet their
electric charges exhibit (in the particular class of models under
consideration here) the same pattern as ordinary quarks. Note that
other classes of 3-4-1 models \cite{key-7} allow for new quarks with
some exotic electric charges such as $\pm4/3$ or $\pm5/3$. However,
the exact colored symmetry $SU(3)_{C}$ of the QCD remains to describe
the strong interaction as a vector-like theory, and its predictions
are not affected at low energies by this enlargement of content. 

The gauge bosons of the electro-weak sector occur in connection with
the standard generators $T_{aL}$ of the $su(4)$ algebra. In this
basis, the gauge fields are $A_{\mu}^{0}$ of $U(1)_{X}$ and $A_{\mu}\in su(4)$,
that is \begin{equation}
A_{\mu}=\frac{1}{\sqrt{2}}\left(\begin{array}{ccccccc}
D_{1\mu} &  & W_{\mu}^{+} &  & K_{\mu}^{+} &  & X_{\mu}^{+}\\
\\W_{\mu}^{-} &  & D_{2\mu} &  & K_{\mu}^{0} &  & X_{\mu}^{0}\\
\\K_{\mu}^{-} &  & K_{\mu}^{0*} &  & D_{3\mu} &  & Y_{\mu}^{0}\\
\\X_{\mu}^{-} &  & X_{\mu}^{0*} &  & Y_{\mu}^{0*} &  & D_{4\mu}\end{array}\right),\label{Eq. 5}\end{equation}
with $D_{1\mu}=A_{\mu}^{3}/\sqrt{2}+A_{\mu}^{8}/\sqrt{6}+A_{\mu}^{15}/\sqrt{12}$,
$D_{2\mu}=-A_{\mu}^{3}/\sqrt{2}+A_{\mu}^{8}/\sqrt{6}+A_{\mu}^{15}/\sqrt{12}$,
$D_{3\mu}=-2A_{\mu}^{8}/\sqrt{6}+A_{\mu}^{15}/\sqrt{12}$, $D_{4\mu}=-3A_{\mu}^{15}/\sqrt{12}$
as diagonal Hermitian bosons. By inspecting Eq.(\ref{Eq. 5}), one
notes that - apart from the charged Weinberg bosons ($W^{\pm}$) -
there are several heavy degrees of freedom, namely two new charged
bosons $K^{\pm}$ and $X^{\pm}$ along with $X^{0}$, $K^{0}$ and
$Y^{0}$ (and their complex conjugated). The diagonal entries provide
us with the neutral physical bosons: the massless photon $A_{\mu}^{em}$
for the electromagnetic interaction and massive $Z_{\mu}$, along
with two heavy neutral bosons $Z_{\mu}^{\prime}$ and $Z_{\mu}^{\prime\prime}$
involved in the neutral currents of the model. 

The scalar quadruplets in order to break the symmetry of the model
stand in the following representations:

\begin{equation}
\phi^{(1)}=\left(\begin{array}{c}
\chi^{0}\\
\chi_{1}^{-}\\
\chi_{2}^{-}\\
\chi_{3}^{-}\end{array}\right)\sim(\mathbf{1,4},-3/2)\label{Eq. 6}\end{equation}

\begin{equation}
\phi^{(i)}=\left(\begin{array}{c}
\rho^{+}\\
\rho_{1}^{0}\\
\rho_{2}^{0}\\
\rho_{3}^{0}\end{array}\right),\quad\left(\begin{array}{c}
\eta^{+}\\
\eta_{1}^{0}\\
\eta_{2}^{0}\\
\eta_{3}^{0}\end{array}\right),\quad\left(\begin{array}{c}
\xi^{+}\\
\xi_{1}^{0}\\
\xi_{2}^{0}\\
\xi_{3}^{0}\end{array}\right)\sim(\mathbf{1,4},1/2)\label{Eq. 7}\end{equation}

The superscripts denote their electric charge in units of $e$ and
$i=2,3,4$. 

The electro-weak sector of the model provides us with two distinct
couplings: $g$ for $SU(4)_{L}$ and $g_{X}$ for $U(1)_{X}$respectively.
Hence, the covariant derivatives read: $D_{\mu}=\partial_{\mu}-i(gA_{\mu}+g_{X}\frac{X}{2}A_{\mu}^{0}){}$.
Evidently, $g$ is the SM coupling of the $SU(2)_{L}$ and $g'$ of
the $U(1)_{Y}$, since $SU(2)_{L}\otimes U(1)_{Y}$ must be a subgroup
of $SU(4)_{L}\otimes U(1)_{X}$. With respect to this subgroup, a
Higgs doublet occurs from $\phi^{(2)}$ - namely $\rho=\left(\begin{array}{c}
\rho^{+}\\
\rho_{1}^{0}\end{array}\right)\sim(\mathbf{1,2},1/2)$ - and two Higgs singlets $\rho_{2}^{0}$, $\rho_{3}^{0}$$\sim(\mathbf{1,1},-1/2)$.
Consequently, there is also $\chi=\left(\begin{array}{c}
\chi^{0}\\
\chi_{1}^{-}\end{array}\right)\sim(\mathbf{1,2},1/2)$ and corresponding singlets $\chi_{2}^{-}$, $\chi_{3}^{-}$$\sim(\mathbf{1,1},-1/2)$.
When the symmetry is spontaneously broken up to the SM electro-weak
group, these two scalar doublets can be seen as the traditional $\phi=\left(\begin{array}{c}
\rho^{+}\\
\rho_{1}^{0}\end{array}\right)\sim(\mathbf{1,2},1)$ of the SM - with the completely decoupled $\rho_{2}^{0}$, $\rho_{3}^{0}$$\sim(\mathbf{1,2},0)$
at this level - and $\tilde{\phi}=i\sigma\phi^{*}=\left(\begin{array}{c}
\chi^{0}\\
\chi_{1}^{-}\end{array}\right)\sim(\mathbf{1,2},-1)$, due to the equivalence $\mathbf{2}\sim\mathbf{2^{*}}$ specific
to $SU(2)$ only. 

Now, the symmetry breaking pattern becomes quite obvious. The four
scalar multiplets in Eqs. (\ref{Eq. 6}) - (\ref{Eq. 7}) break the
symmetry of the model in three steps to the residual one, namely to
the electromagnetic $U(1)_{em}$:

\[
\begin{array}{ccc}
SU(3)_{C}\otimes SU(4)_{L}\otimes U(1)_{X} & \underrightarrow{V'} & SU(3)_{C}\otimes SU(3)_{L}\otimes U(1)_{X^{'}}\\
\\ & \underrightarrow{V} & SU(3)_{C}\otimes SU(2)_{L}\otimes U(1)_{Y}\\
\\ & \underrightarrow{v+v'} & SU(3)_{C}\otimes U(1)_{em}\end{array}\]
by developing the vacuum expectation values (vev): $<\phi^{(1)}>=\left(\begin{array}{cccc}
v^{\prime} & 0 & 0 & 0\end{array}\right)^{T}$ , $<\phi^{(2)}>=\left(\begin{array}{cccc}
0 & v & 0 & 0\end{array}\right)^{T}$, $<\phi^{(3)}>=\left(\begin{array}{cccc}
0 & 0 & V & 0\end{array}\right)^{T}$, $<\phi^{(4)}>=\left(\begin{array}{cccc}
0 & 0 & 0 & V^{\prime}\end{array}\right)^{T}$. A reasonable alignment is assumed here $v\cong v^{\prime}\ll V\cong V^{\prime}$
in order to get rapid and simpler estimations. Evidently, $v$, $v'=174$GeV
is the SM electro-weak breaking scale, while the new scales $V$,
$V^{\prime}$ are specific to this 3-4-1 model.

In the symmetry breaking limit one can easily obtain the couplings
match, namely:

\[
\frac{1}{g'^{2}}=\frac{1}{2g^{2}}+\frac{1}{g_{X}^{2}}\]

This relation leads straightforwardly to the 

\begin{equation}
\frac{g_{X}^{2}}{g^{2}}=\frac{\sin^{2}\theta_{W}(m_{X})}{1-\frac{3}{2}\sin^{2}\theta_{W}(m_{X})}\label{Eq. 8}\end{equation}

When applying the renormalization group procedure to Eq. (\ref{Eq. 30})
in the limit $\sin^{2}\theta_{W}=\sin^{2}\theta_{W}(m_{Z})$ it results
that the unification scale $m_{X}$ is not sensitive to $\alpha_{X}(m_{Z})$
which remains $\alpha_{X}(m_{Z})\sim10^{-2}$ even for $m_{X}$up
to $10^{16}GeV$ (while $\alpha(m_{Z})\sim1/128$) . 

Of special interest is the boson mass spectrum. It can be computed
as:

\begin{equation}
m_{W^{\pm}}^{2}=\frac{1}{2}g^{2}\left(v^{2}+v'^{2}\right)\quad,\quad m_{Y^{0}(Y^{0*})}^{2}=\frac{1}{2}g^{2}\left(V^{2}+V'^{2}\right)\quad,\label{Eq. 9}\end{equation}

\begin{equation}
m_{K^{\pm}}^{2}=\frac{1}{2}g^{2}\left(V^{2}+v'^{2}\right)\quad,\quad m_{K^{0}(K^{0*})}^{2}=\frac{1}{2}g^{2}\left(V^{2}+v^{2}\right)\quad,\label{Eq. 10}\end{equation}

\begin{equation}
m_{X^{\pm}}^{2}=\frac{1}{2}g^{2}\left(V'^{2}+v'^{2}\right)\quad,\quad m_{X^{0}(X^{0*})}^{2}=\frac{1}{2}g^{2}\left(V'^{2}+v^{2}\right)\quad,\label{Eq. 11}\end{equation}

\begin{equation}
m_{Z}^{2}=\frac{m_{W^{\pm}}^{2}}{\cos^{2}\theta_{W}}\quad,\quad M_{Z',Z''}^{2}=\frac{1}{2}g^{2}\left(\begin{array}{ccc}
\frac{V^{2}+V'^{2}}{4} &  & \frac{V^{2}-V'^{2}}{2}\\
\\\frac{V^{2}-V'^{2}}{2} &  & V^{2}+V'^{2}\end{array}\right).\label{Eq. 12}\end{equation}

In our approximation $V\cong V^{\prime}$, one can easily diagonalize
the above matrix and get: $m_{Z'}^{2}=\frac{1}{4}g^{2}V^{2}$ and
$m_{Z''}^{2}=g^{2}V^{2}$. $Z_{\mu}^{\prime\prime}$ couples only
with exotic fermions having no interaction with ordinary particles
and thus totally decoupling from the low energy phenomenology, while
$Z^{\prime}$ mixes with the SM $Z$ and exhibits a mass (\cite{key-7,key-8})
lower bounded by $2TeV$ (or greater), in close dependence on which
one of the three generations of quarks transforms differently from
the other two. The restrictions on the $Z^{\prime}$ neutral boson
come from plugging into this model the data supplied by atomic parity
violation experiments and certain processes in the meson systems $D^{0}-\bar{D^{0}}$,
$B_{s}^{0}-\bar{B_{s}^{0}}$, $B_{d}^{0}-\bar{B_{d}^{0}}$, $K^{0}-\bar{K^{0}}$
(with their corresponding mixings in the FCNC). However, these theoretical
bounds \cite{key-7,key-8} are consistent with recent experimental
observations \cite{key-1} that suggest a lower bound around $1TeV$
for $Z^{\prime}$. 

In order to ensure the consistency with the SM phenomenology, the
heavy particles of the 3-4-1 symmetry must not compromise the precision
tests of the SM, particularly the oblique corrections must remain
unaffected. That is, the new heavy quarks, new heavy bosons and the
extra scalar fields must give negligible contributions to the oblique
parameters $U$, $S$, $T$, namely to the vacuum polarization amplitudes
of the $W$ and $Z$ bosons. Being completely decoupled from low energy
physics, $Z_{\mu}^{\prime\prime}$ does not contribute to these oblique
corrections. Also, $Y^{0}(Y^{0*})$ does not contribute as long it
is a singlet under $SU(2)$. The $SU(2)$ doublets $(K^{+},K^{0})$and
$(X^{+},X^{0})$ (their complex conjugates) does not alter the one-loop
calculations due to their degenerate masses (see Eqs. (\ref{Eq. 10})
- (\ref{Eq. 11})). There remains to be investigated only the contribution
of some scalar doublets such as $\left(\begin{array}{c}
\eta^{+}\\
\eta_{1}^{0}\end{array}\right)$ and $\left(\begin{array}{c}
\xi^{+}\\
\xi_{1}^{0}\end{array}\right)$ whose couplings are subject to a proper tuning. However, these detailed
calculations exceeds the aim of this paper. 

All the fermion masses are generated through Higgs mechanism by scalar
particles interacting with fermion fields. In this connection, the
large split of breaking scales works efficiently in preserving the
SM phenomenology. Ordinary quarks and leptons acquire their masses
at the SM scale $v$ while exotic quarks at the new $V,V'$ scales. 

A vast amount of theoretical research has been accomplished in this
field due to several striking features that make such SM-extensions
quite appealing and much valuable despite the fact that they claim
for a plethora of new particles. We count some of their assets. (i)
First of all, the generation number in the fermion sector seems to
get its explanation (absent in the SM, where simply an \emph{ad hoc}
triplication of the first generation is performed). In order to cancel
the chiral anomalies - this time by an interplay among families -
the number of generations must be divisible by the number of colors
$N_{C}=3$. This leads to exactly $3$ generations if one assumes
the asymptotic freedom condition from QCD that limits them to no more
than 5. (ii) Contrary to SM, these models supply a natural framework
for charge quantisation (Doff and Pisano, \cite{key-7}). (iii) The
strong CP problem can be elegantly solved due to a natural existence
of Peccei-Quinn symmetry (Pal, Montero, Sanchez-Vega \cite{key-5}).
(iv) It offers a suitable framework for implementing the little Higgs
mechanism (Kong \cite{key-7} ). (v) All SM neutral currents and masses
are identically recovered. (vi) All new particles acquire their masses
from the high scales $V$, $V'$ so they do not interfere with the
SM phenomenology at low energies supplied by present facilities. (vii)
If the third generation of quarks is the one transforming differently,
that accounts naturally for the unbalancing heaviness of the top quark
. 

We mention that our work focuses on a particular model from the 3-4-1
class of models, namely the one corresponding in the systematic classification
accomplished by Ponce and Sanchez \cite{key-9} to $b=c=1$, Model
A. Its rich phenomenology of such models can be found and compared
in Refs. \cite{key-7,key-8}.

\section{Lepton masses }

The Yukawa sector of any gauge model is set up to supply fermion masses
(consequently the SSB). There are introduced certain Yukawa coefficients
(matrix $h$) that couple left-handed and right-handed fermion fields.
We write down the most general combinations allowed by the gauge symmetry. 

\begin{equation}
\mathcal{L}_{Y}^{lept}=h_{ii}^{l}\bar{f}_{iL}\phi^{(1)+}l_{iR}+\frac{1}{\Lambda}\bar{f}_{iL}\left(h_{ij}^{M}S_{R}^{+}f_{jL}^{c}+h_{ij}^{D}S_{D}^{+}f_{jL}^{c}+h_{ij}^{D}S_{D}^{\prime+}f_{jL}^{c}\right)+h.c.\label{Eq. 13}\end{equation}
where $l_{L}=e_{L},\mu_{L},\tau_{L}$. $S$ matrices are defined as
follows $S_{R}=(\phi^{(3)}\otimes\phi^{(4)}+\phi^{(4)}\otimes\phi^{(3)})\sim(\mathbf{1},\mathbf{10},1)$,
$S_{D}=(\phi^{(2)}\otimes\phi^{(3)}+\phi^{(3)}\otimes\phi^{(2)})\sim(\mathbf{1},\mathbf{10},1)$,
$S_{D}^{\prime}=(\phi^{(2)}\otimes\phi^{(4)}+\phi^{(4)}\otimes\phi^{(2)})\sim(\mathbf{1},\mathbf{10},1)$
with $h_{ij}^{M}$, $h_{ij}^{D}$ as thegeneric Yukawa matrices for
Dirac and Majorana terms. Evidently, this is the basis where the charged
leptons are already diagonal, so $h_{ij}^{l}=0$. Up to this point
we proved that the electro-weak gauge symmetry $SU(4)_{L}\otimes U(1)_{X}$
allows for a natural implementation of neutrino masses via dimension-five
effective operators plus the Yukawa couplings, usually subject to
certain extra assumptions to overcome their arbitrariness.

In order to restrict ourselves to a simpler version the entries in
the generic neutrino matrix are the same regardless their nature,
namely $h_{ij}^{M}=h_{ij}^{D}$. Therefore, the Yukawa Lagrangian
becomes in the case at hand here:

\begin{equation}
\mathcal{L}_{Y}^{lept}=h_{ii}^{l}\bar{f}_{iL}\phi^{(1)+}l_{iR}+h_{ij}\left[\frac{1}{\Lambda}\bar{f}_{iL}\left(S_{R}^{+}f_{jL}^{c}+S_{D}^{+}f_{jL}^{c}+S_{D}^{\prime+}f_{jL}^{c}\right)\right]+h.c.\label{Eq. 14}\end{equation}

In concrete expressions below the Yukawa coefficients will be denoted
in order $A=h_{ee}$, $B=h_{\mu\mu}$, $C=h_{\tau\tau}$, $D=h_{e\mu}$,
$D'=h_{\mu e}$ $E=h_{e\tau}$, $E'=h_{\tau e}$, $F=h_{\mu\tau}$,
$F'=h_{\tau\mu}$. 

It is natural to consider that the positions $3$ and $4$ in each
lepton quadruplet are precisely $N=\nu_{R}$ and $N^{\prime}=\nu_{R}^{c}$.
As long as they are sterile with respect to $Z$ and exhibit indistinguishable
couplings to the new $Z^{\prime}$ and $Z^{\prime\prime}$ bosons
(see for instance Ref. \cite{key-8}) This assumption leads straightforwardly
to the following identification:

\begin{equation}
\mathcal{L}_{Y}^{\nu}=\mathcal{L}_{Y}^{R}+\mathcal{L}_{Y}^{D}+\mathcal{L}_{Y}^{D^{\prime}}.\label{Eq. 15}\end{equation}

By inspecting Eq.(\ref{Eq. 13}) one can easily identify

\begin{equation}
m(e)=h_{e}^{l}v,\quad m(\mu)=h_{\mu}^{l}v\,,\quad m(\tau)=h_{\tau}^{l}v\,.\label{Eq. 16}\end{equation}
 since we work in a basis where the charged lepton sector is diagonal.

Taking into consideration the field theory mass formulas for respectively
Dirac ($\mathcal{L}_{Y}^{D}=-m_{D}\bar{\psi^{c}}\psi+H.c.$) and Majorana
($\mathcal{L}_{Y}^{M}=-\frac{1}{2}m_{M}\bar{\psi^{c}}\psi+H.c.$)
terms, along with the above vev alignment and the normal seesaw (\ref{Eq. 1}),
one gets the mass matrices:

\begin{equation}
M_{D}=h\frac{vV}{\Lambda},\label{Eq. 17}\end{equation}

\begin{equation}
M_{R}=\left(h+h^{T}\right)\frac{V^{2}}{\Lambda}.\label{Eq. 18}\end{equation}

From Eq. (\ref{Eq. 2}), in the flavor basis the Majorana terms for
left-handed and right-handed neutrinos can be read 

\begin{equation}
M(\nu_{L})=\left[h^{T}\left(h+h^{T}\right)^{-1}h\right]\frac{v^{2}}{\Lambda},\label{Eq. 19}\end{equation}

\begin{equation}
M(\nu_{R})=\left(h+h^{T}\right)\frac{V^{2}}{\Lambda}.\label{Eq. 20}\end{equation}

By diagonalizing Eq. (\ref{Eq. 18}) with a proper $U_{R}$ one obtains
those matrices in the mass basis. First step is:

\begin{equation}
\widehat{\left(h+h^{T}\right)}=Diag\left(r_{1},r_{2},r_{3}\right)=U_{R}^{T}\left(h+h^{T}\right)U_{R}\label{Eq. 21}\end{equation}

Now, inserting this result in Eq. (\ref{Eq. 19}) one computes

\begin{equation}
M(\nu_{L})=\left[h^{T}U_{R}\widehat{\left(h+h^{T}\right)^{-1}}U_{R}^{T}h\right]\frac{v^{2}}{\Lambda}\label{Eq. 22}\end{equation}

which can be diagonalized as $\widehat{M_{L}}=U^{T}M(\nu_{L})U$ ,
namely:

\begin{equation}
\widehat{M_{L}}=\left[U^{T}h^{T}U_{R}\widehat{\left(h+h^{T}\right)^{-1}}U_{R}^{T}hU\right]\frac{v^{2}}{\Lambda}\label{Eq. 23}\end{equation}

The physical neutrino masses can be computed via Eq.(\ref{Eq. 23})
if we consider their mixing (for details, see the reviews in Ref.
\cite{key-10}). The unitary mixing matrix $U$ ($U^{+}U=1$) links
the gauge-flavor basis to the physical basis of massive neutrinos:

\begin{equation}
\nu_{\alpha L}(x)=\sum_{i=1}^{3}U_{\alpha i}\nu_{iL}(x)\label{Eq. 24}\end{equation}
where $\alpha=e,\mu,\nu$ (corresponding to neutrino gauge eigenstates),
and $i=1,2,3$ (corresponding to massive physical neutrinos with masses
$m_{i}$). The mixing matrix $U_{PMNS}$ (Pontecorvo-Maki-Nakagawa-Sakata)
has in the standard parametrization the form: 

\begin{equation}
U_{PMNS}=\left(\begin{array}{ccc}
c_{12}c_{13} & s_{12}c_{13} & s_{13}e^{-i\delta}\\
-s_{12}c_{23}-c_{12}s_{13}s_{23}e^{i\delta} & c_{12}c_{23}-s_{12}s_{13}s_{23}e^{i\delta} & c_{13}s_{23}\\
s_{12}s_{23}-c_{12}s_{13}c_{23}e^{i\delta} & -c_{12}s_{23}-s_{12}s_{13}c_{23}e^{i\delta} & c_{13}c_{23}\end{array}\right)\label{Eq. 25}\end{equation}
where the notations $\sin\theta_{23}=s_{23}$, $\sin\theta_{12}=s_{12}$,
$\sin\theta_{13}=s_{13}$, $\cos\theta_{23}=c_{23}$, $\cos\theta_{12}=c_{12}$,
$\cos\theta_{13}=c_{13}$ stand for the mixing angles and $\delta$
is the Dirac CP phase (with phenomenological meaning). Usually, to
this one a diagonal Majorana phase matrix $P=Diag\left(1,e^{i\alpha},e^{i\beta}\right)$
is sticked, though it can be absorbed by redefining fields. The standard
identification leads to solar angle - $\theta_{12}$, atmospheric
angle - $\theta_{23}$, reactor angle - $\theta_{13}$. Several different
patterns for matrix \ref{Eq. 25} has been considered in the literature.
The most appealing approach stemming from the seminal work of Harrison
Perkins, Scott \cite{key-11}, the so called {}``tri-bi-maximal''
ansatz is largely invoked when the PMNS matrix is analysed and its
phenomenological consequences are worked out in different models.
A possible alternative to it followed the {}``bi-maximal'' line
and was developed in Refs. \cite{key-12}. These particular textures
- in good agreement with data - are often unfolded by enforcing certain
discrete flavor symmetries on $M$.

The global data \cite{key-1} regarding neutrino oscillations impose
certain restrictions, namely $\sin^{2}\theta_{12}\simeq0.3$, $\sin^{2}\theta_{23}\simeq0.5$,
and a small reactor angle $\theta_{13}$(probably near zero), along
with the mass splittings $\Delta m_{12}^{2}\simeq7.6\times10^{-5}eV^{2}$
and $\Delta m_{23}^{2}\simeq2.4\times10^{-3}eV^{2}$.

\section{Plausible scenario and numerical estimates}

In this section we prove that our construction is not a mere theoretical
device. On contrary it can work very well when it comes to confronting
the experimental data. We don't claim to make a general analysis here
and get general predictions, but just apply the above procedure to
a very particular scenario to get a quick fit with the data. The results
make it obvious that our approach can be further developed and employed
in investigating the neutrino sector.

\subsection{Left-handed physical neutrinos}

Since $U_{R}$ is not restricted on observational ground, one can
assume for the sake of simplicity a suitable scenario in which $h_{ji}=-h_{ij}$
- namely $D'=-D$, $E'=-E$, $F'=-F$ - that leads to $U_{R}=I$,
and hence:

\begin{equation}
\left(h+h^{T}\right)=Diag(2A,2B,2C)\,,\quad\left(h+h^{T}\right)^{-1}=Diag(\frac{1}{2A},\frac{1}{2B},\frac{1}{2C})\label{Eq. 26}\end{equation}

and consequently:

\begin{equation}
M(\nu_{L})\simeq\frac{1}{2}\left(\begin{array}{ccc}
A+\frac{D{}^{2}}{B}+\frac{E{}^{2}}{C} & \frac{EF}{C} & -\frac{DF}{B}\\
\\\frac{EF}{C} & B+\frac{D{}^{2}}{A}+\frac{F{}^{2}}{C} & \frac{DE}{A}\\
\\-\frac{DF}{B} & \frac{DE}{A} & C+\frac{E{}^{2}}{A}+\frac{F{}^{2}}{B}\end{array}\right)\frac{v^{2}}{\Lambda}\label{Eq. 27}\end{equation}

Bearing in mind that $Trace$ is independent of the basis we work
in, so that $TrM(\nu_{L})=\sum_{i}m_{iL}$, one obtains: 

\begin{equation}
TrM(\nu_{L})=\frac{v^{2}}{2\Lambda}\left(A+\frac{D{}^{2}}{B}+\frac{E{}^{2}}{C}+B+\frac{D{}^{2}}{A}+\frac{F{}^{2}}{C}+C+\frac{E{}^{2}}{A}+\frac{F{}^{2}}{B}\right)\label{Eq. 28}\end{equation}

If all the Yulawa couplings are in the same range with the coupling
of the charged $\tau$lepton (at most comparable but no greater than
it) then one obtains an upper bound for the sum of the individual
neutrino masses. This is:

\begin{equation}
\sum_{i}m_{iL}\leq\frac{9}{2}m(\tau)\frac{v}{\Lambda}\label{Eq. 29}\end{equation}
As long as the experimental evidence imposes an upper bound on the
range of left-handed neutrino masses (a few $eV$), one can estimate
the cit-off energies up to which this model is valid. If $v=174GeV$
and $m(\tau)=1777MeV$, then $\Lambda\geq1.4\times10^{12}GeV$ which
evidently is an intermediate level between SM and GUT energies.

\subsection{Mass hierarchy and mixing angles}

In order to get some rough estimates of the mass spectrum and fit
properly the mixings between the three species of left-handed neutrinos,
let's consider a simple setting: $A\simeq-0.371\times10^{-4}$, $B\simeq C\simeq-0.5383\times10^{-3}$,
$D\simeq E\simeq0.9034i\times10^{-3}$, $F\simeq-0.17266i\times10^{-2}$,
where $i^{2}=-1$ the complex unity.

Under these circumstances, the left-handed neutrino mass matrix (\ref{Eq. 27})
becomes:

\begin{equation}
M(\nu_{L})\simeq\left(\begin{array}{ccc}
0.150 & 0.145 & -0.145\\
0.145 & 1.350 & 1.100\\
-0.145 & 1.100 & 1.350\end{array}\right)\times10^{-2}\frac{v^{2}}{\Lambda}\label{Eq. 30}\end{equation}
so that it can be roughly diagonalized by the unitary matrix

\begin{equation}
U=\left(\begin{array}{ccc}
0.831 & 0.556 & 0\\
-0.393 & 0.587 & 0.707\\
0.393 & -0.587 & 0.707\end{array}\right)\label{Eq. 31}\end{equation}
corresponding to the experimentally observed values $\sin^{2}\theta_{12}\simeq0.31$,
$\sin^{2}\theta_{23}\simeq0.5$, and$\sin^{2}\theta_{13}\simeq0$. 

The mass spectrum comes out in this particular case in a normal hierarchy,
namely

\begin{equation}
\left|m_{1}\right|\simeq0.00216eV,\quad m_{2}\simeq0.08866eV\,,\quad m_{3}\simeq0.52983eV\,.\label{Eq. 32}\end{equation}

The mass splitting ratio yields $r_{\Delta}\simeq0.03$ in good agreement
with experimental data and so the mass squared splittings are. 

Of course, a further work could consider a non-zero CP phase violation
and a small but non-zero reactor angle so that from this stage one
can perform a more accurate calculus and take into consideration a
plethora of scenarios once the solar ($\theta_{12}$) and atmospheric
angles ($\theta_{23}$) are firmly established. All these scenarios,
of course, cax lead to many different $M(\nu_{L})$ matrices and,
finally, even to a different mass hierarchy. Our results here are
nothing but the proof that our method can work and is not claimed
to be a general analytical analysis (which will be performed in a
future work).

\subsection{Right-handed sterile neutrinos}

Now, a few words about the right-handed neutrinos in our approach
framework. They acquire (Eq.(\ref{Eq. 26})) the following masses
in the scenario at hand: 

\begin{equation}
\sum_{i}m_{iR}=\frac{2V^{2}}{\Lambda}\left(A+B+C\right)\label{Eq. 33}\end{equation}

According to the assumption in Sec. 4.1, this can be approximated
as: 

\begin{equation}
\sum_{i}m_{iR}\simeq6m(\tau)\frac{V^{2}}{v\Lambda}\label{Eq. 34}\end{equation}
or equivalently: 

\begin{equation}
\sum_{i}m_{iR}\simeq\frac{4}{3}\left(\frac{V}{v}\right)^{2}\sum_{i}m_{iL}\label{Eq. 35}\end{equation}
which, as expected, is not affected by the cut-off $\Lambda$. If
$V$ is not very high - say around $1-10TeV$ - so that its new physics
is testable at LHC, the so called sterile neutrinos develop masses
in the sub-$keV$ range. The stability of these three species of right-handed
neutrinos could recommend they as good candidates for the warm component
of the dark matter \cite{key-14}. Indeed, their only interactions
with left-handed neutrinos are mediated by heavy bosons $K$ and $X$
(see Eq.(\ref{Eq. 5})) via such couplings as $\frac{g}{\sqrt{2}}$$\overline{(\nu_{R})^{c}}\gamma^{\mu}\nu_{L}X_{\mu}^{0}$
and $\frac{g}{\sqrt{2}}$$\overline{(\nu_{R})}\gamma^{\mu}\nu_{L}K_{\mu}^{0}$.
When calculating the width for a kinematically allowed decay $\nu_{R}\longrightarrow\nu_{R}^{\prime}\nu_{L}\nu_{L}^{\prime}$,
the expression $\Gamma=G_{F}^{2}m_{\nu_{R}}^{5}m_{W}^{4}/192\pi^{3}m_{X,K}^{4}$
satisfactorily supplies the order of magnitude. The life times $\tau=\Gamma^{-1}\sim10^{27}s$,
provided the fact that both bosons are in $TeV$ mass region. This
result is by far greater than the estimated age of the Universe $t_{0}\simeq4.3\times10^{17}s$
according to \cite{key-15}, so the right-handed neutrinos as WIMP
particles have a safe behavior from the stability viewpoint. A deeper
investigation of the cosmological and astrophysical implications  exceeds
the scope of this letter and will be performed in a future work.

\section{Summary}

In this paper we worked out the canonical seesaw mechanism in the
framework of electro-weak \textbf{$SU(4)_{L}\otimes U(1)_{Y}$} extension
of the SM, by simply constructing dimension-five effective operators
as suitable direct products among scalar quadruplets existing in the
model. Then we made use of the same generic Yukawa matrix for Dirac
and Majorana couplings in order to obtain the plausible mass spectrum
for the left-handed neutrinos. It came out in the range $10^{-3}-10^{-1}eV$.
For this purpose, an intermediate scale between SM and GUT was set
by the cut-off $\Lambda\simeq1.4\times10^{12}GeV$. The right-handed
partners develop masses in dependence on the high breaking scale $V$
of the model. If the latter one ranges in $1-10TeV$ domain, then
the right-handed neutrinos can be seen as plausible warm Dark Matter
candidates with mass in $10^{-1}keV$. 

The advantage of our method over other approaches is that it exploits
in a natural way all the ingredients supplied by the 3-4-1 model itself,
without enlarging the scalar sector of the model, without invoking
any new particle in the lepton sector or a fine-tuning procedure.
Of course, a more detailed analysis can further be performed based
on different hypotheses regarding the Yukawa couplings that can be
subject to a new and appropriate flavor symmetry.


\begin{thebibliography}{10}
\bibitem{key-1}Particle Data Group (K. Nakamura \emph{et al.}), \emph{J. Phys. G}
\textbf{37}, 075021 (2010).
\bibitem{key-2}SuperKamiokande Collab., \emph{Phys. Rev. Lett.} \textbf{81}, 1562
(1998); \emph{Phys. Rev. D} \textbf{71}, 112005 (2005); K2K Collab.,
\emph{Phys. Rev. Lett.} \textbf{96}, 181801 (2006); SNO Collab., \emph{Phys.
Rev. Lett.} \textbf{92}, 181301 (2004); KamKAND Collab., \emph{Phys.
Rev. Lett.} \textbf{94}, 081801 (2005); LSND Collab., \emph{Phys.
Rev. D} \textbf{64}, 112007 (2001). 
\bibitem{key-3}R. N. Mohapatra and G. Senjanovic, \emph{Phys. Rev. Lett.} \textbf{44},
912 (1980); T. Yanagida in \emph{Proceedings of the Workshop on the
Unified Theory and Baryon Number in the Universe} (ed. O Sawada and
A. Sugamoto, KEK 1979); M. Gell-Mann, P. Ramond and R. Slansky in
\emph{Supergravity} (ed. P. van Neuwenhuizen and D. Z. Freedman, North
Holland, 1979); J. Schechter and J. W. F. Valle, \emph{Phys. Rev.
D} \textbf{22}, 2227 (1980).
\bibitem{key-4}S. Weinberg, \emph{Phys. Rev. Lett.} \textbf{43}, 1566 (1979); F.
Wilczek and A. Zee, \emph{Phys. Rev. Lett.} \textbf{43}, 1571 (1979);
E. Ma and U. Sarkar, \emph{Phys. Rev. Lett.} \textbf{80}, 5716 (1998);
E. Ma, \emph{Phys. Rev. Lett.} \textbf{81}, 1171 (1998); E. Ma, \emph{Phys.
Rev. Lett.} \textbf{86}, 2502 (2001); W. Grimus, L. Lavoura and B.
Radovcic, \emph{Phys. Lett}. \emph{B} \textbf{674}, 117 (2009); K.
S. Babu, S. Nandi and Z. Tavartkiladze, \emph{Phys. Rev. D} \textbf{80},
071702 (2009); F. Bonnet, D. Hernandez, T. Ota and W. Winter, \emph{JHEP}
\textbf{0910}, 076 (2009); I. Picek and B. Radovcic, \emph{Phys. Lett.}
\emph{B} \textbf{687}, 338 (2010).
\bibitem{key-5}P. H. Frampton, \emph{Phys. Rev. Lett.} \textbf{69}, 2889 (1992);
F. Pisano and V. Pleitez, \emph{Phys. Rev.} \emph{D} \textbf{46},
410 (1992); P. Pal, \emph{Phys. Rev. D} \textbf{52}, 1659 (1995);
H. N. Long, \emph{Phys. Rev.} \emph{D} \textbf{53}, 437 (1996); A.
Palcu, \emph{Mod. Phys. Lett. A} \textbf{21}, 1203 (2006); J. C. Montero
and B. L. Sanchez-Vega, \emph{Phys. Rev. D} \textbf{84}, 055019 (2011). 
\bibitem{key-6}J. C. Montero, C. A de S. Pires and V. Pleitez, \emph{Phys. Rev. D}
\textbf{65}, 095001 (2002); A. G. Dias, C. A de S. Pires and P. S.
Rodrigues da Silva , \emph{Phys. Lett. B} \textbf{628}, 85 (2005);
A. Palcu \emph{Mod. Phys. Lett. A} \textbf{21}, 2591 (2006); D. Cogollo,
H. Diniz, C. A de S. Pires and P. S. Rodrigues da Silva , \emph{Eur.
Phys. J. C} \textbf{58}, 455 (2008); P. V. Dong and H. N. Long , \emph{Phys.
Rev. D} \textbf{77}, 0573002 (2008); D. Cogollo, H. Diniz and C. A
de S. Pires, \emph{Phys. Lett. B} \textbf{677}, 338 (2009); D. Cogollo,
H. Diniz and C. A de S. Pires, \emph{Phys. Lett. B} \textbf{687},
400 (2010). 
\bibitem{key-7}R. Foot, H. N. Long and T. A. Tran, \emph{Phys. Rev. D} \textbf{50},
R34 (1994); F. Pisano and V. Pleitez, \emph{Phys. Rev. D} \textbf{51},
3865 (1995); A. Doff and F. Pisano, \emph{Mod. Phys. Lett. A} \textbf{14},
1133 (1999); A. Doff and F. Pisano, \emph{Phys. Rev. D} \textbf{63},
097903 (2001); Fayyazuddin and Riazuddin, \emph{JHEP} \textbf{0412},
013 (2004); O. C. W. Kong, \emph{Phys. Rev. D} \textbf{70}, 075021
(2004); L. A. Sanchez, F. A Perez and W. A. Ponce, \emph{Eur. Phys.
J C} \textbf{35}, 259 (2004); L. A. Sanchez, L. A. Wills-Toro and
J. I. Zuluaga, \emph{Phys. Rev. D} \textbf{77}, 035008 (2008); J.
L. Nisperuza and L. A. Sanchez, \emph{Phys. Rev. D} \textbf{80}, 035003
(2009); S. Villada and L. A. Sanchez, \emph{J. Phys. G} \textbf{36},
115002 (2009); A. Palcu, \emph{Mod. Phys. Lett. A} \textbf{24}, 2589
(2009); A. Jaramillo and L. A. Sanchez, \emph{Phys. Rev. D} \textbf{84},
115001 (2011). 
\bibitem{key-8}W. A. Ponce, D. A. Gutierrez and L. A. Sanchez, \emph{Phys. Rev. D}
\textbf{69}, 055007 (2004); Riazuddin and Fayyazuddin, \emph{Eur.
Phys. J C} \textbf{56}, 389 (2008); A. Palcu, \emph{Int. J. Mod. Phys.
A} \textbf{24}, 4923 (2009). 
\bibitem{key-9}W. A. Ponce and L. A. Sanchez, \emph{Mod. Phys. Lett. A} \textbf{22},
435 (2007); A. Palcu, \emph{Mod. Phys. Lett. A} \textbf{24}, 1247
(2009). 
\bibitem{key-10}A. Strumia and F. Visani, \emph{Neutrino masses and mixing and...},
arXiv: hep-ph/0606054v3; R. N. Mohapatra and A. Y. Smirnov, \emph{Ann.
Rev. Nucl. Part. Sci.} \textbf{56}, 569 (2006); .Zhi-zhong Xing, \emph{Nucl.
Phys. B (Proc. Suppl.)} \textbf{203-204}, 82 (2010). 
\bibitem{key-11}P. F. Harrison, D. H. Perkins and W.G Scott, \emph{Phys Lett B} \textbf{530},
167 (2002); P. F. Harrison and W. G. Scott, \emph{Phys. Lett. B} \textbf{535},
163 (2002); Zhi-zhong Xing, \emph{Phys. Lett. B} \textbf{533}, 85
(2002); E. Ma, \emph{Phys. Rev. D} \textbf{66}, 117301 (2002); C.
Low and R. R. Volkas, \emph{Phys. Rev. D} \textbf{68}, 033007 (2002);
F. Plentinger and W. Rodejohann, \emph{Phys Lett. B} \textbf{625},
264 (2005); Nan Li and Bo-Qiang Ma, \emph{Phys. Rev. D}\textbf{71},
017302 (2005); W. Grimus and L. Lavoura, \emph{JHEP} \textbf{0601},
018 (2006); G. Altarelli and F. Feruglio, \emph{Nucl. Phys. B} \textbf{741},
215 (2006); I. de Medeiros Varzielas, S. F. King and G. G. Ross, \emph{Phys.
Lett. B} \textbf{644}, 153 (2007); I. de Medeiros Varzielas, S. F.
King and G. G. Ross, \emph{Phys. Lett. B} \textbf{648}, 201 (2007);
Y. Koide, J. Phys. G \textbf{34}, 1653 (2007); E. Ma, \emph{Europhys.
Lett.} \textbf{79}, 61001 (2007); P. H. Frampton and T. W Kephart,
\emph{JHEP} \textbf{0709}, 110 (2007); C. H. Albright and W. Rodejohann,
\emph{Phys. Lett. B} \textbf{665}, 378 (2008); S. F. King, \emph{Phys.
Lett. B} \textbf{659}, 244 (2008); C. D. Carone and R. F. Lebed, \emph{Phys.
Rev D} \textbf{80}, 117301 (2009); M. Hirsch, S. Morisi and J. W.
F. Valle, \emph{Phys. Lett. B} \textbf{679}, 454 (2009); S. Goswani,
S. T. Petcov, S. Ray and W. Rodejohann, \emph{Phys. Rev. D} \textbf{80},
053013 (2009); G. Altarelli and D. Meloni, \emph{J. Phys. G} \textbf{36},
085005 (2009); S. F. King, \emph{Phys. Lett. B} \textbf{675}, 347
(2009); W. Grimus and L. Lavoura, \emph{J. Phys. G} \textbf{36}, 115007
(2009); Gui-Jun Ding, \emph{Nucl. Phys. B} \textbf{827}, 82 (2010);
M. Abbas and A. Y. Smirnov, \emph{Phys. Rev. D} \textbf{82}, 013008
(2010); D. Meloni, F. Plentinger and W. Winter, \emph{Phys. Lett.
B} \textbf{699}, 354 (2011); Y. Shimizu and R. Takahashi, \emph{Europhys.
Lett.} \textbf{93}, 61001 (2011); Zhen-hua Zhao, \emph{Phys. Lett.
B} \textbf{701}, 609 (2011); S. Antusch, S. F. King, C. Luhn and M.
Spinrath, \emph{Nucl. Phys. B} \textbf{850}, 477 (2011); X.-G. He
and A. Zee, \emph{Phys. Rev. D} \textbf{84}, 053004 (2011) ; T. Araki,
\emph{Phys. Rev D} \textbf{84}, 037301(2011) .
\bibitem{key-12}V. D. Barger, S. Pakvasa, T. J. Weiler and K. Whisnant, \emph{Phys.
Lett. B} \textbf{437}, 107 (1998); A.J. Baltz, A. Scharff and M. Goldhaber,
\emph{Phys. Rev. Lett.} \textbf{81}, 5730 (1998); G. Altarelli and
F. Feruglio, \emph{Phys. Lett. B} \textbf{439}, 112 (1998); G. Altarelli
and F. Feruglio, \emph{JHEP} \textbf{9811}, 021 (1998); S. Davidson
and S. F. King, \emph{Phys. Lett. B} \textbf{445}, 191 (1998); D.
V. Ahluvalia, \emph{Mod. Phys. Lett. A} \textbf{13}, 2249 (1998);
R. N. Mohapatra and S. Nussinov, \emph{Phys. Rev. D} \textbf{66},
113002 (1999); C. Giunti, \emph{Phys. Rev. D} \textbf{59}, 177301
(1999); C. Jarlskog, M. Matsuda, S. Skadhauge and M. Tanimoto, \emph{Phys.
Lett. B} \textbf{449}, 240 (1999); M. Jezabek and Y. Sumino, \emph{Phys.
Lett. B} \textbf{457}, 139 (1999); Yue-Liang Wu, \emph{Eur. Phys.
J C} \textbf{10} 491 (1999); W. G. Scott, \emph{Nucl. Phys. Proc.
Suppl}. \textbf{85}, 177 (2000); Zhi-zhong Xing, \emph{Phys. Rev.
D} \textbf{61}, 057301 (2000); T. Kitabayashi and M. Yasue, arXiv:
hep-ph/006040; R. N. Mohapatra, \emph{Phys. Rev. D} \textbf{64}, 091301
(2001); K. Choi, E. J. Chun, K. Hwang and W. Y. Song, \emph{Phys.
Rev. D} \textbf{64}, 113013 (2001); M. Jezabek, \emph{Nucl. Phys.
Proc. Suppl}. \textbf{100}, 276 (2001); K. S. Babu and R. N. Mohapatra,
\emph{Phys. Lett. B} \textbf{532}, 77 (2002); C. Giunti and M. Tanimoto,
\emph{Phys. Rev. D} \textbf{66}, 053013 (2002); W. Rodejohann, \emph{Phys.
Rev. D} \textbf{70}, 073010 (2004); S. K. Kang and C. S. Lim, \emph{Phys.
Lett. B} \textbf{634}, 520 (2006); G. Altarelli and F. Feruglio, \emph{JHEP}
\textbf{0905}, 020 (2009); C. H. Albright, A. Duek and W. Rodejohann,
\emph{Eur Phys. J. C} \textbf{70}, 1099 (2010).
\bibitem{key-13}S. Morisi, K. M. Patel and E. Peinado, \emph{Phys. Rev. D} \textbf{84},
053002 (2011); H. Fritsch, Z.-z. Xing and S. Zhou, \emph{JHEP} \textbf{1109}
083 (2011); D. Meloni, \emph{JHEP} \textbf{1110}, 010 (2011); H. Ishimori
and T. Kobayashi, \emph{JHEP} \textbf{1110}, 082 (2011); D. Marzocca,
S. T. Petcov, A. Romanino and M. Spinrath, \emph{JHEP} \textbf{1111}
009 (2011); S. Zhou, \emph{Phys. Lett. B} \textbf{704}, 291 (2011);
H. Zhang and S. Zhou, \emph{Phys. Lett. B} \textbf{704}, 296 (2011);
W. Rodejohann, H. Zhang and S. Zhou, \emph{Nucl. Phys.} \emph{B} \textbf{855},
592 (2012); G. J. Ding and D. Meloni, \emph{Nucl. Phys. B} \textbf{855},
21 (2012); S. Antusch, S. F. King, C. Luhn and M. Spinrath, \emph{Nucl.
Phys. B} \textbf{856}, 328 (2012).
\bibitem{key-14}P.L. Biermann and A. Kusenko, \emph{Phys. Rev. Lett} \textbf{96},
091301 (2006); T. Asaka, M. Laine and M. Shaposhnikov, \emph{JHEP}
\textbf{0701}, 091 (2007); Wan-lei Guo, \emph{Phys. Rev D} \textbf{77},
033005 (2008); G. B. Gelmini, E. Osoba and S. P. Ruiz, \emph{Phys.
Rev. D} \textbf{81}, 063529 (2010); F. Bezrukov, H. Hettmansperger
and M. Lindner, \emph{Phys. Rev. D} \textbf{81}, 085032 (2010); S.
Ando and A. Kusenko, \emph{Phys. Rev. D} \textbf{81}, 113006 (2010);
M. Lindner, A. Merle and V. Niro, \emph{JCAP} \textbf{1101}, 034 (2011);
A. Merle and V. Niro, \emph{JCAP} \textbf{1107}, 023 (2011); C. R.
Watson, Zh. Li, and N. K. Polley, arXiv: 11114217.
\bibitem{key-15}D. N. Spergel et al. (WMAP Collab.), \emph{Astrophys. J. Suppl.} \textbf{148},
175 (2003) 175. \end{thebibliography}
\end{document}